\pdfoutput=1
\documentclass[twocolumn,showpacs,amssymb,floatfix,aps]{revtex4}

\usepackage{graphicx}
\usepackage{dcolumn}
\usepackage{bm}
\usepackage{epsfig}

\begin{document}
CERN-PH-TH/2010-241
\preprint{CERN-PH-TH/2010-241}
\title{
QED confronts the radius of the proton
}
\author{A. De R\'ujula${}^{a,b,c}$}
\affiliation{  \vspace{3mm}
${}^a$ Instituto de F\'isica Te\'orica (UAM/CSIC), Univ. Aut\'onoma de Madrid, Madrid, and 
CIEMAT, Madrid, Spain,\\
${}^b$ Physics Dept., Boston University, Boston, MA 02215,\\
${}^c$Physics Department, CERN, CH 1211 Geneva 23, Switzerland}

\date{\today}

\begin{abstract}

Recent results on muonic hydrogen [1] and the ones compiled by CODATA on ordinary hydrogen and $ep$-scattering [2] are $5\sigma$ away from each other. Two reasons justify a further look at this subject: 1) One of the approximations used in [1] is not valid for muonic hydrogen. This amounts to a shift of the proton's radius by $\sim 3$ of the standard deviations of [1], in the ``right" direction of data-reconciliation.
In field-theory terms, the error is a mismatch of renormalization scales. Once corrected, the proton radius ``runs", much as the QCD coupling ``constant" does. 2) The result of [1] requires a choice of the ``third Zemach moment". Its published independent determination is based on an analysis with a $p$-value --the probability of obtaining data with equal or lesser agreement with the adopted (fit form-factor) hypothesis-- of $3.92\times 10^{-12}$. In this sense, this quantity is not empirically known. Its value would regulate the level of ``tension" between muonic- and ordinary-hydrogen results, currently {\it at most} $\sim 4\sigma$. There is no tension between the results of [1] and the proton radius determined with help of the analyticity of its form factors.

\end{abstract}


\maketitle

\section{Introduction}

The results of a measurement by Pohl et al.~\cite{Pohl}
of the Lamb shift in muonic hydrogen 
and those compiled by CODATA on ordinary hydrogen and $ep$-scattering 
\cite{CODATA} are $\sim 5 \,\sigma$ away from each other.
The authors of \cite{Pohl} conclude ``Our result
implies that either the Rydberg constant has to be shifted by
2110 kHz/c (4.9 standard deviations), or the calculations of the
QED effects in atomic hydrogen or muonic hydrogen atoms are
insufficient.'' I discuss why the second option is part of the
resolution of the apparent conundrum, but not all of it.

It is intrepid \cite{rpADR} to use a model of the proton  --in \cite{Pohl}, a dipole form-factor--
to challenge very well established physics --such as QED \cite{Pohl,Flowers}.
But this is not the only bone of contention:

One of the approximations used
in the theory of ordinary or muonic hydrogen involves the lepton's wave
function at the origin. The approximation is sufficiently good for the former atom,
but not the latter. The required correction can be rephrased by having an $r_p$ that
runs, in the same sense as $\alpha_s$ --the fine structure constant of QCD-- does.
The modification results in a $\sim 3\,\sigma(\mu H)$ shift of the extracted 
central value of $r_p$,
in the direction of reducing the ``tension" between experimental results.
This correction depends on the model of the proton's charge distribution,
but the model-dependence is a small correction to a moderate correction.
These issues are discussed in detail in Sections III and IV.

The current way to extract  $r_p$ from $ep$ scattering data
involves an extrapolation to a momentum transfer
${\mathbf q^2}=0$, the point from which $\langle r_p^2 \rangle$ is inferred.
This extrapolation covers a two-orders of magnitude larger hiatus
than the one relevant to muonic hydrogen;
the model-dependence is correspondingly larger. 
The extrapolated object is
a form-factor fit to data gathered above 
$|{\mathbf q}^2|={\cal{O}}(m_\pi^2)$, a domain where there is still ``structure",
relative to, e.g.~a dipole form factor \cite{Mainz}.

The extraction of $r_p$ from $ep$ data has severe statistical
problems, mentioned in the abstract and discussed in Section V. 
One way to reappraise this issue is to take new, very precise data \cite{Mainz},
see also Section V. The difficulties associated with these analyses are
shared by the measurement of the other relevant quantity: 
the ``third Zemach moment", as discussed in Section VI and VII.

Sections VIII is a discussion of the experimental and theoretical results.
 Section IX contains my conclusions. 

\section{The issue}

Former precise measurements of $r_p$ had two origins. One is
mainly based on the theory \cite{HTheory} and observations  \cite{Hexps} of  
hydrogen. The 
result, compiled in CODATA \cite{CODATA}, is
\begin{equation}
\sqrt{\langle r_p^2\rangle}\rm (CODATA)=0.8768 \pm 0.0069\; \rm fm.
\label{rCODATA}
\end{equation}
The second type of measurement is based on the theory and observations
\cite{Sick, Sick2} of very low-energy electron-proton scattering. 
An analysis of the world data as of a few years ago  yielded \cite{Sick}:
\begin{equation}
\sqrt{\langle r_p^2\rangle}(ep)=0.895 \pm 0.018\; \rm fm
\label{rep}
\end{equation}

A recent $ep$-scattering experiment \cite{Mainz} results in:
 \begin{equation}
\sqrt{\langle r_p^2\rangle}\rm (A1)=
0.879(5)_{stat}(4)_{syst}(2)_{model}(4)_{group} \;\rm fm,
\label{rA1}
\end{equation}
whose various subindices will be clarified anon.

The proton's charge distribution, $\rho_p(r)$,
 is related to the non-relativistic limit of the 
electric form-factor, $G_E$, by the Fourier transformation 
\begin{equation}
G_E(-{\mathbf q}^2)=\int d^3 r \,\rho_p({\mathbf r})\, e^{-i\,\vec q\, \vec r}
\simeq 1-{{\mathbf q}^2\over 6}\,{\langle r_p^2\rangle}+\, ...\, ,
\label{F}
\end{equation}
which also serves to define ${\langle r_p^2\rangle}$ in proportion to the
${\mathbf q}^2$-derivative of the form factor at ${\mathbf q}^2=0$.

The most precise relevant measurement to date is that of the
 $\rm 2P_{3/2}^{F=2}\to 2S_{1/2}^{F=1}$ Lamb shift in the $\mu p$ atom, 
\cite{Pohl}:
\begin{equation}
L_{\rm exp} = 206.2949 \pm 0.0032\;\rm meV.
\label{Lexp}
\end{equation}
In meV units for energy and fermi units for the radii, the predicted value 
\cite{LymanTH}
is of the form
\begin{eqnarray}
L^{\rm th}\left[\langle r_p^2\rangle,\langle r_p^3 \rangle_{(2)}\right]&=&\nonumber\\
209.9779(49)&-&5.2262\, \langle r_p^2\rangle +0.00913 \,\langle r_p^3 \rangle_{(2)}
\label{Lth}
\end{eqnarray}
The first two coefficients are best  estimates of many contributions 
 while the third stems from the $n=2$ value of an addend
  \cite{FriarSick,HTheory}
 \begin{equation}
 \Delta E_3(n,l)= {\alpha^5\over 3\,n^3}\,m_r^4 \,
 \delta_{l0}\,\langle r_p^3 \rangle_{(2)},
 \label{DeltaE3}
 \end{equation}
 proportional to the third Zemach moment
\begin{equation}
\langle r_p^3 \rangle_{(2)}\equiv\int d^3 r_1 d^3 r_2\,\rho(r_1)\rho(r_2)
\vert {\mathbf r}_1-{\mathbf r}_2\vert^3
\end{equation}

For a specific model of $\rho_p(r)$ or its corresponding $G_E(-{\mathbf q}^2)$,
the two $r$-moments in Eq.~(\ref{Lth}) are related. For instance, for a dipole
form factor
\begin{equation}
\left[\langle r^3\rangle_{(2)}\right]^2={(3675/ 256)} \left[\langle r^2\rangle\right]^3
\label{dipoleapprox}
\end{equation}
while for a single pole 
$\left[\langle r^3\rangle_{(2)}\right]^2= {(50/ 3)} \left[\langle r^2\rangle\right]^3$.

The authors of \cite{Pohl} use the dipole relation of Eq.~(\ref{dipoleapprox})
 in Eq.~(\ref{Lth})
to convert Eq.~(\ref{Lexp}), into an impressively accurate
\begin{equation}
\sqrt{\langle r_p^2\rangle}(\mu\rm H)= 0.84184\pm 0.00067 \; fm
\label{rPohl}
\end{equation}

The value of $r_p(\mu H)$ in Eq.~(\ref{rPohl})
differs by $\sim 3\,\sigma(ep)$ from Eq.~(\ref{rep}),
$5.0\,\sigma\rm(CODATA)$ from Eq.~(\ref{rCODATA}), and a bit 
more from Eq.~(\ref{rA1}). The standard deviations of these last three $r_p$
determinations are much bigger than the ones in Eq.~(\ref{rPohl}).
Thus, they essentially
determine the significance of the ``distance" to the latter result.

\section{Insufficiently-good approximations}

Let $\ell$ stand for $e$, $\mu$ and
let $m_r\equiv m_\ell \,m_p/(m_\ell + m_p)$ be the reduced mass.
In an $\ell p$ atom the dominant contribution (99.45\% of the total
for $\ell =\mu$) to the coefficient
of the ${\langle r_p^2\rangle}$ term in Eq.~(\ref{Lth}) is the familiar:
\begin{equation}
-\Delta E_{\{n=2;\,l =0\}}^{\rm FS}=
{2\pi\alpha\over 3}{\langle r_p^2\rangle}\vert \Psi_{{2,0}}(0)\vert^2
=
{\alpha^4\over 12}\,m_r^3 {\langle r_p^2\rangle}
\label{oldr2shift}
\end{equation}
Recall that, in writing Eq.~(\ref{oldr2shift}), the Fourier transform
($V=-4\,\pi\,\alpha/ {\mathbf q}^2$)  of a
Coulomb potential ($V=-\alpha/r$) has been modified by the expression
in the rhs of Eq.~(\ref{F}) to obtain an additive term, $\propto \delta(\vec r)$,
resulting in the ``0" in the argument of the atom's wave function $\Psi$.

Even for $\ell = \mu$, the Bohr radius $a_{\rm B}=1/(\alpha\, m_r)$ is orders of magnitude
larger than $r_p$, apparently justifying the consuetudinary
approximation used in the last paragraph, which results in the $\Psi(0)$ factor.
But the precision of the measurement in Eq.~(\ref{Lexp}) and its allegedly consequent
 Eq.~(\ref{rPohl}) is so unprecedented, that the approximation must be revisited,
 as I proceed to do.
 
Consider a dipole form-factor
$G_E(-{\mathbf q}^4)\equiv m_d^4/(m_d^2+{\mathbf q}^2)^2$, for which
$m_d^2=12/{\langle r_p^2\rangle}$. Repeat the analysis leading to 
Eq.~(\ref{oldr2shift}), this time without making the
approximation of Eq.~(\ref{F}). The result is
\begin{eqnarray}
&&\!\!\!\!\!\!\!-\Delta E_{\{n=2;\,l =0\}}^{\rm FS}=
{\alpha\over 4\,a_{\rm B}}{1+3\, m_d\,a_{\rm B}+4 \,m_d^3\,a_{\rm B}^3
 \over (1+m\,a_{\rm B})^5}=\nonumber\\
&&\!\!\!\!\!\!\!\!\!\! {\alpha^4\over 12} m_r^3 {\langle r_p^2\rangle}\!
\left(1-5\,\alpha\,m_r\,\sqrt{\langle r_p^2 \rangle \over 12}
+{\cal{O}}\left[{1\over (m_d\,a_{\rm B})^{2}}\right]\right)
\label{dipole}
\end{eqnarray}
Naturally, the leading term coincides with Eq.~(\ref{oldr2shift}). 
The first order correction to $r_p$, estimated by entering the 
$r_p^2$ value of Eq.~(\ref{rPohl}) amounts to 0.42\%.
This may  look tiny. But it increases
the value of $r_p$, extracted as in Eq.~(\ref{rPohl}), by  
$2.7\,\sigma(\mu H)$. This modification of the central value of $r_p$,
though also insufficient by itself, is in the 
direction of reconciling the body of experimental results.

It is also instructive to consider, for the nonce, a single-pole form-factor 
$G_E(-{\mathbf q}^2)\equiv m^2/(m^2+{\mathbf q}^2)$, for which
$m^2=6/{\langle r_p^2\rangle}$.
The result is
\begin{eqnarray}
&&\!\!\!\!\!\!\!-\Delta E_{\{n=2;\,l =0\}}^{\rm FS}=
{\alpha\over 4\,a_{\rm B}}{1+2\, m^2\,a_{\rm B}^2 \over (1+m\,a_{\rm B})^4}=
\nonumber\\
&&\!\!\!\!\!\!\!\!{\alpha^4\over 12} m_r^3 {\langle r_p^2\rangle}\!
\left(1-4\,m_r\,\alpha\,\sqrt{\langle r_p^2\rangle \over 6}
+{\cal{O}}\left[{1\over (m\,a_{\rm B})^{2}}\right]\right)
\label{pole}
\end{eqnarray}
This correction amounts to 0.48\%, or $3\,\sigma(\mu H)$.
Substitute $m_\mu$ for $m_e$ to conclude the obvious:
for ordinary hydrogen and the precision of the
corresponding observations, the corrections of 
Eqs.~(\ref{pole},\ref{dipole})
are negligible.

We have learned that, at the level of accuracy of the $\mu p$ experiment,
the evaluation of the ${\langle r_p^2\rangle}$ term in Eq.~(\ref{Lexp})
is not only delicate; it is also model-dependent. This is because of
the inevitable extrapolation to ${\mathbf q}^2=0$, where the radius is defined.
We shall see that in the extraction of information from $ep$ experiments, 
for which the extrapolation covers a two-orders of magnitude larger gap,
the model-dependence is correspondingly larger.

\section{A running ${\langle r_p^2\rangle}$}

The ``atomic" subtleties discussed in the previous section are very familiar in QCD.
To discuss the simplest analogy,
consider the total cross section for $e^+e^-$ annihilation
into hadrons, above or in-between
quark thresholds. It is of the form $\sigma(Q^2)\!\propto\!(1/Q^2)(1+\alpha_s/\pi+...)$.
For the approximation to be correct at all $Q^2$, $\alpha_s$ must ``run",
that is, be $Q^2$-dependent in a specific way.  

In the next simplest example,
the $n$-th moment of a (non-singlet) proton structure function
--analogous to ${\langle r_p^n\rangle}$-- if evaluated at two $Q^2$ values,
differs by a multiplicative factor: to leading order, the ratio 
$\alpha_s(Q^2)/\alpha_s(Q'^2)$ to a specific anomalous dimension, $d_n$.

In a field theory, an expression like Eq.~(\ref{oldr2shift}), containing a $\Psi(r)$ and 
an ${\langle r_p^2\rangle}$ referring to two different scales, would be
a ``mistmatch of renormalization points". To
 correct it, one must  evaluate $\Psi$
at the correct distance scale (as in the previous paragraph)
 or let ${\langle r_p^2\rangle}$ run. For a chosen form-factor
this statement can be made precise, e.g.~even the ``$5/\sqrt{12}$" in Eq.~(\ref{dipole})
has some meaning. 

The proton is not probed by the orbiting muon at $r=0$,
or by momenta with equal weights in the range $(0,\infty)$ in
the Fourier transform of $\delta(\vec r)$. It is only probed by momenta 
ranging from $|{\mathbf q}|={\cal{O}}(\alpha\,m_r)$ up to
$|{\mathbf q}|\sim m_d/4$, the proper ``ultraviolet" scale. To use the mid
expression in Eq.~(\ref{oldr2shift}) at a consistent distance scale,
$\vec r=0$, one may, for a dipole, {\it define} 
${\langle r_p^2\rangle}\vert_{(\alpha m_r,m_d)}\equiv m_d^2/12$,
keep the term $|\Psi(0)|^2$ and substitute 
${\langle r_p^2\rangle}$ by a running radius
\begin{equation}
{\langle r_p^2\rangle}\vert_{(0,\infty)}\simeq
{{\langle r_p^2\rangle}\vert_{(\alpha m_r,m_d)}\over
1+5\,\alpha\,m_r \,\sqrt{\langle r_p^2\rangle\vert_{(\alpha m_r,m_d)} /12}},
\label{runningr}
\end{equation}
very reminiscent of the expression for $\alpha_s$ in QCD.
It is the lhs of Eq.~(\ref{runningr}) that is needed
to extract the slope of the form factor at ${\mathbf q}^2=0$,
as in the rhs of Eq.~(\ref{F}).

The dipole form factor is not foreign to QCD.
The understanding of the relatively high-$Q^2$ physics summarized by
a dipole approximation, and the deviations thereof \cite{A,GT} --as
well as the related first measurement of $\Lambda_{\rm QCD}$ \cite{A}-- were discussed
immediately after the discovery of QCD's asymptotic freedom \cite{PGW}.

\section{The extraction of ${\langle r_p^2\rangle}$ from $ep$ scattering data}

\begin{figure}[hbt!]
\hspace{-.4cm}
\includegraphics[width=0.5\textwidth]{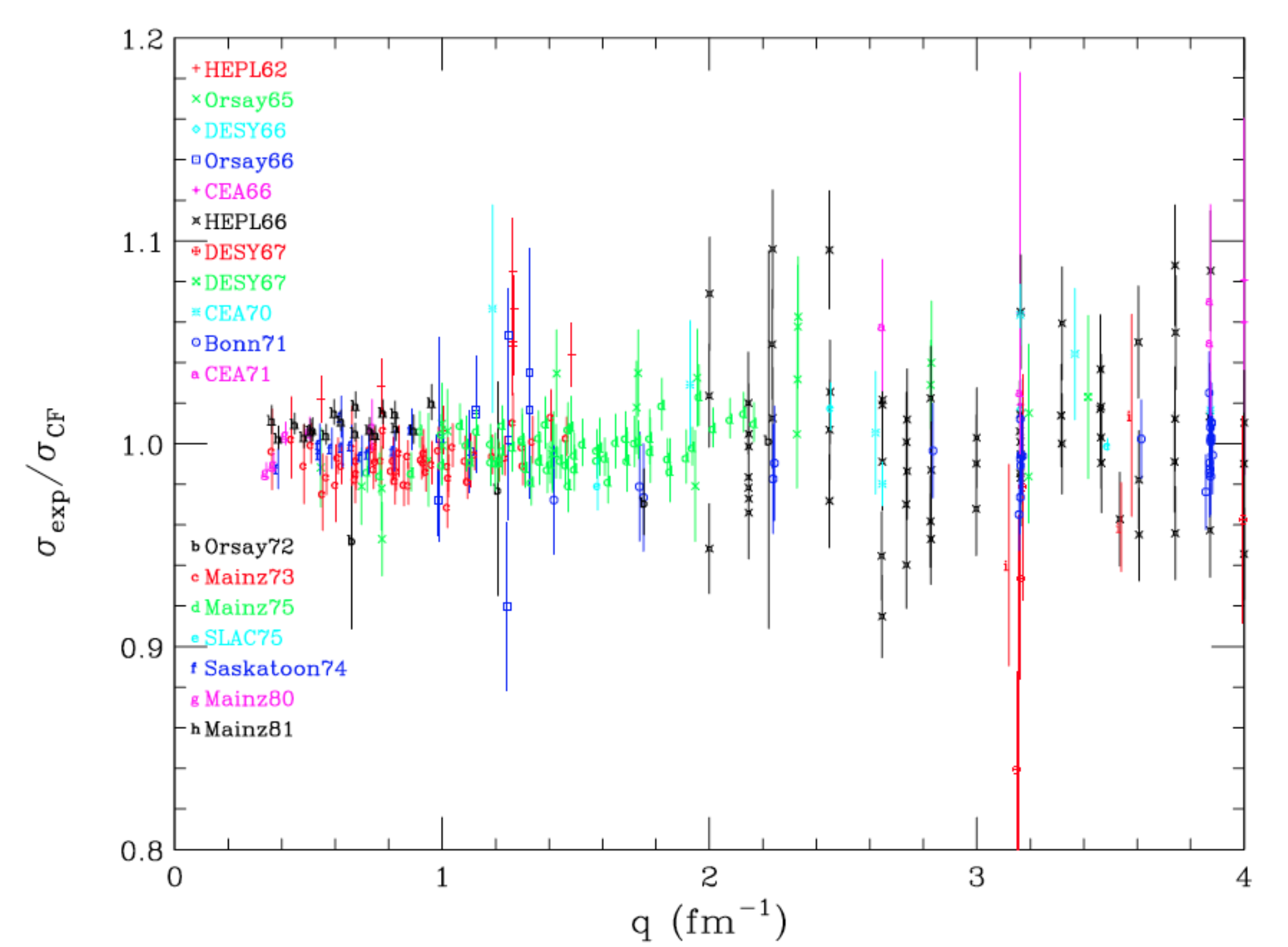}
\caption{
Low-$|{\mathbf q}|$ data, compiled and analysed in \cite{Sick, FriarSick}.
  \label{fig:Sick}}
\end{figure}

The Lyman-shift result quoted in Eq.~(\ref{rPohl}) is $\sim\! 3\,\sigma(ep)$ away from
the $ep$-scattering result of Eq.~(\ref{rep}) . This is not a severe problem. 
A look at the data, reproduced in Fig.~\ref{fig:Sick}, on which the latter result is based,
indicates that the problem if even less severe.
What is shown in the figure are data available in 2003, normalized to 
a 5-parameter continued-fraction expansion of $G_E(-{\mathbf q}^2)$ \cite{Sick}. 
The fit's result is ${\chi}^2/n_{\rm dof}\!\simeq\!1.652$, or, more explicitly, $\chi^2\,\simeq\,512$ for 
$n_{\rm dof}\!=\!310$ degrees of freedom. 

It may be useful to recall that the $p$-value of a data-set relative to a given
assumption or fit --in this case the specified continued fraction--
is the probability of obtaining
data at least as incompatible with the hypothesis as the data actually observed.
Let $f(\chi^2,n_{\rm dof})$ be the $\chi^2$ probability distribution function. 
Let $\Gamma(a,b)\,[\Gamma(b)]$ be the incomplete [ordinary] gamma 
function. Then
\begin{equation}
p(\chi^2,n_{\rm dof})=\int_{\chi^2}^{\infty}f(z,n_{\rm dof})\,dz=
{\Gamma(n_{\rm dof}/2,\chi^2/2)\over \Gamma(n_{\rm dof}/2)}
\label{pvalue}
\end{equation}
and $p(512,310)\!\simeq\! 3.92\times 10^{-12}$, i.e.~the quality of the
fit in \cite{Sick} is not ``quite good".
It is possible \cite{Sick} to reduce this behemoth disagreement  by 
adding quadratically 3\% to the Stanford error bars
(to obtain $p(370,310)\!\simeq\!0.011$),
or by a norm change of 1\% of \cite{Simon}... [which] would decrease $\chi^2$
by 60 (resulting in $p(452,310)\!\simeq\!2.38\times 10^{-7}$).
Modifying the data 
is not necessarily a universally accepted procedure, or so would the corrected
experimentalists opine.

It is also possible to draw sensible-looking curves through the data that, 
in their slope at ${\mathbf q}^2=0$, differ from a straight horizontal line
in Fig.~\ref{fig:Sick} by one or more of the $\sigma$'s in Eq.~(\ref{rep}). 
The fact that the data points are very scattered
is an unavoidable problem. One way to reconsider the issue is to take new
and very precise data. 

\begin{figure}[hbt!]
\hspace{-.4cm}
\includegraphics[width=0.5\textwidth]{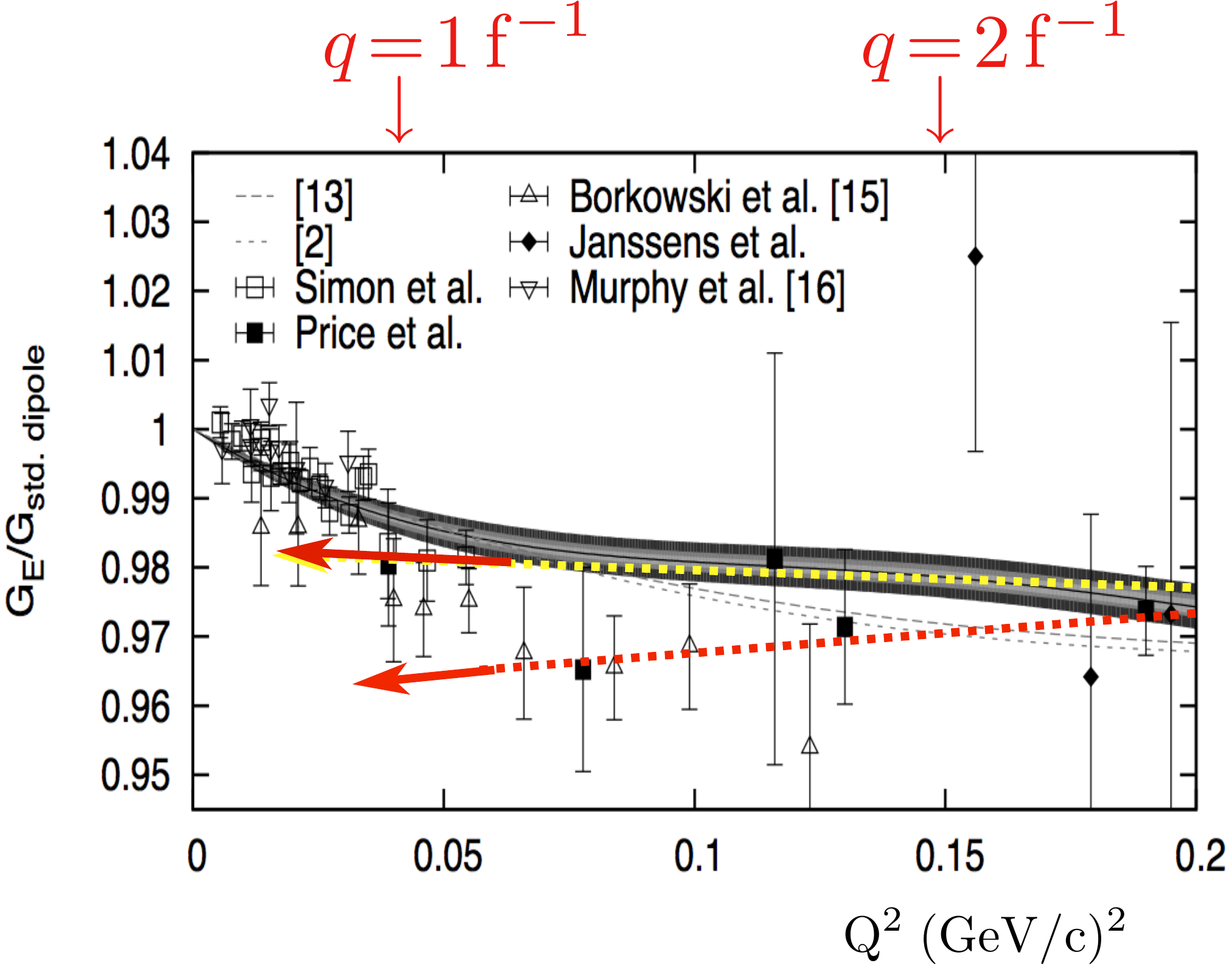}
\caption{The lowest-energy Mainz data \cite{Mainz}. The dark trumpet encompasses
estimated theoretical and systematic uncertainties and $\pm 1\sigma$ statistical
errors. Barred points are results of other experiments. The dotted lines are by-eye
fits to data in the domain $\Delta Q^2=(\sim 0.06,0.2)$ (GeV/c)$^2$, for each of the
sets of data. Their arrowy continuations are meant to illustrate how difficult it
may be to extract  from such an
extrapolation the slope at the lowest-$Q^2$ measured point.
\label{fig:Bernauer}}
\end{figure}

Such data exist \cite{Mainz} and are partially reproduced in Fig.~\ref{fig:Bernauer}.
The paper contains many relevant commentaries.
One of them is: ``The structure at small $Q^2$ seen in $G_E$ and
$G_M$ corresponds to the scale of the pion of about 
$Q^2\approx m_\pi^2\approx 0.02$ (GeV/c$^2$) and may be indicative
of the influence of the pion clowd." The most apposite remarks
in \cite{Mainz} concern the extraction of the results:


Two types of ``flexible" models are considered in \cite{Mainz}: fits to polynomials
and spline fits. The $G_E$ results are
\begin{eqnarray}
\sqrt{\langle r_p^2\rangle}\rm (spline) =\rm
0.875(5)_{stat}(4)_{syst}(2)_{model} \;\rm fm,&&\nonumber\\
\sqrt{\langle r_p^2\rangle}\rm (polynom) =\rm
0.883(5)_{stat}(5)_{syst}(3)_{model} \;\rm fm &&
\label{rA1double}
\end{eqnarray}
``Despite detailed studies the cause of the difference could not be found.
Therefore, we give as the final result the average of the two values with
an additional uncertainty of half of the difference"\cite{Mainz}: the outcome 
quoted in Eq.~(\ref{rA1}).
Whether the fits' uncertainties are thus correctly estimated is
 debatable, but this is not the main point.

The crux of the matter is that
the procedure in \cite{Mainz}  illustrates how,
even for sets of ``flexible" fits, the result is significantly set-dependent.
The reason is simple: two analytic (or piece-wise analytic)
functions arbitrarily close to
each other in a given interval, say $\Delta Q^2=(Q^2_{\min},\infty)$, can be
arbitrarily different in their continuation to another interval, such
as $\Delta Q^2=(0,Q^2_{\min})$.

The data itself could be used to study the model-dependence of the
extracted value of $\langle r_p^2 \rangle$. Suppose
that one fits the data in the interval $\Delta Q^2=(\sim 0.06,0.2)$ 
GeV$^2$ of 
Fig.~\ref{fig:Bernauer} and extrapolates
to the lowest-${\mathbf q}^2$ point at which there is still data and
the $G_E$ slope is measured. This is analogous to extrapolating
to ${\mathbf q}^2=0$, except in that the answer is known. A look at
Fig.~\ref{fig:Bernauer} suffices to conclude that the result is likely to be significantly
wrong.

The ``less flexible" models used to analyse the Mainz data have 
$\chi^2/n_{\rm dof}\approx 1.16$ to 1.29 for $n_{\rm dof}\approx 1400$ \cite{Mainz}.
The corresponding $p$-values range from $2.69\times 10^{-5}$ to
$9.07\times 10^{-13}$. The most flexible ones have $\chi^2/n_{\rm dof}\approx 1.14$,
or $p=1.88\times 10^{-4}$, that is, they are far from being flexible enough to describe
the data. There seems to be a general tendency to forget that the quality of
a fit is a function of two variables, not of their ratio, and that for large fixed
$n_{\rm dof}$ the dependence of  the fit's quality on $\chi^2$ is an inordinately
sharp function around $\chi^2=n_{\rm dof}$, suffice it to plot Eq.~(\ref{pvalue})
to convince oneself.

\section{The extraction of $\langle r_p^3 \rangle_{(2)}$ from $ep$ scattering data}

The result for the third Zemach moment is  \cite{FriarSick}
\begin{equation}
[\langle r_p^3 \rangle_{(2)}]^{1/3}=(1.394\pm 0.022)\;\rm fm,
\label{r3FS}
\end{equation}
based on the data in Fig.~\ref{fig:Sick}. Based on the same data, the result for 
$\langle r_p^2 \rangle$ is that of Eq.~(\ref{rep}).
To discuss the ``third" moment, it is useful to
write it in an alternative form:
\begin{eqnarray}
\langle r_p^3 \rangle_{(2)}&\approxeq&
\int_{0}^{\infty}d{|\mathbf q}|\;I({\mathbf q}^2)\nonumber\\
\nonumber\\
I({\mathbf q}^2)&\equiv& {48\over \pi\,{\mathbf q}^4}\left[G_E^2({\mathbf q}^2)-1
+{{\mathbf q}^2\over 3}\,{\langle r_p^2\rangle}\right]
\label{r3alter}
\end{eqnarray}
Notice that $I({\mathbf q}^2)$ tends to a constant as ${\mathbf q}^2\to 0$.

\begin{figure}[hbt!]
\hspace{-.4cm}
\includegraphics[width=0.5\textwidth]{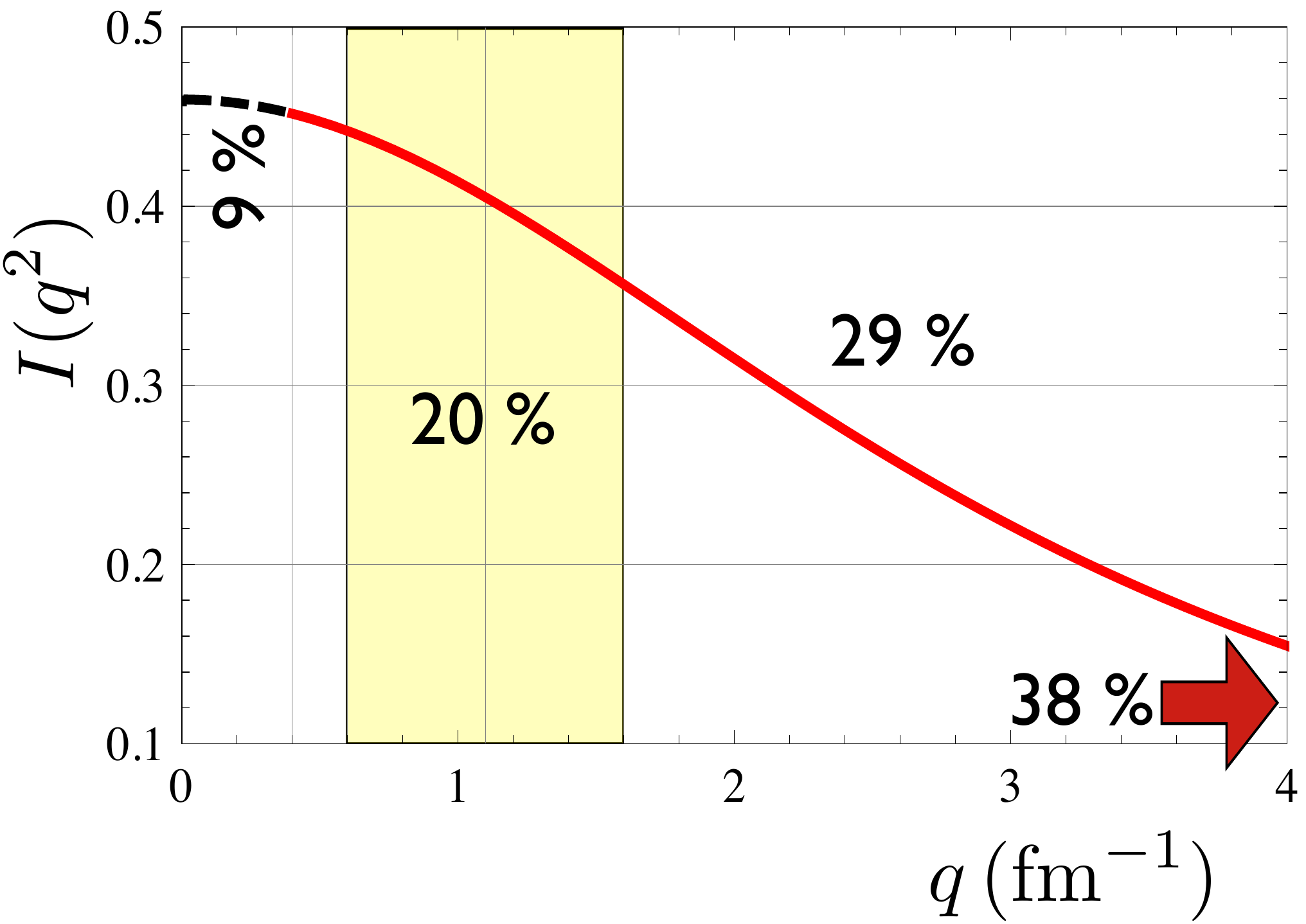}
\caption{
The third Zemach integrand of Eq.~(\ref{r3alter}), illustrated for the
usual dipole and shown in the same domain as that of
Fig.~(\ref{fig:Sick}). The shaded band is at $\Delta q=1.1\pm 0.5$
fm$^{-1}$. Data in the dashed extrapolation do not exist. The contributions
to the moment from various intervals are discussed in the text.
  \label{fig:3Zemach}}
\end{figure}

The usual dipole form factor ($m_d^2=0.71$ GeV$^2$)
is a good-enough approximation for the forthcoming discussion.
The shape of $I({\mathbf q}^2)$ is shown in Fig.~\ref{fig:3Zemach}.
Normalized to the total integral, the fractions of the integral in
various relevant intervals $\Delta q=(a,b)$ fm$^{-1}$ are the following:
\begin{eqnarray}
&&(0,0.4)\to 0.09\,;\;\;
(0.6,1.6)\to 0.20\,;\;\;\nonumber\\
&&(1.6,4)\to 0.29\,;\;\;
(4,\infty)\to 0.38\,.
\label{intervals}
\end{eqnarray}

It is stated in \cite{FriarSick} that ``Sensitivity studies have shown
that the main contribution to the integral comes from the region
$\Delta q=(0.6,1.6)$ fm$^{-1}$ where the data base for electron-proton scattering
is very good." But the contribution is significantly
larger in the range $\Delta q\times\rm fm=(1.6,4)$ where the data are particularly bad,
see Fig.~\ref{fig:Sick}. The contribution if the range $(4,\infty)$
is even larger. The contribution in the range $(0,\sim 0.4)$, where
there are simply no data, is not at all negligible. In other words,
given the results in Eq.~(\ref{intervals}),
the quoted statement must refer to the error estimate,
not to the central value. And the basis for the deduced
central value on $\langle r_p^3 \rangle_{(2)}$
is still the fit in Fig.~\ref{fig:Sick}, whose $p$-value I have quoted.

It is concluded in another study \cite{CM} of $I({\mathbf q}^2)$ that 
``a large third Zemach moment can only occur if
$\langle r_p^4 \rangle$ is also large". This is
unobjectionable, though ``large" means relative to the
expectation from a dipole form-factor.

\section{How to extract $r_p$ moments from $ep$ scattering data?}

We have seen that the values and errors obtained in the literature
for $\langle r_p^2 \rangle$ and $\langle r_p^3 \rangle_{(2)}$ are 
not credible. Since the various moments are highly correlated, another
pertinent question is: how to draw, in the 
$(\langle r_p^2 \rangle,\langle r_p^3 \rangle_{(2)})$ plane, trustable,
model-independent
contour plots of given significance? A procedure might be the following:

\noindent (1) Normalize the data to a fixed model, as 
in Figs.~\ref{fig:Sick},\ref{fig:Bernauer}.

\noindent (2) Study modifications, relative to the model, with a complete
set of orthogonal functions, e.g.~a discrete Fourier basis for the
complete data interval, a function of a variable, such as $\log {\mathbf q}^2$,
chosen to emphasize the most relevant, low-${\mathbf q}^2$, domain.

\noindent (3) Let the results fix the needed flexibility, i.e.~cut
the Fourier series at the term for which $\chi^2\leqslant n_{\rm dof}$
($p\leqslant 0.5$).

\noindent (4) Sidestep an extrapolation to $\mathbf q\to 0$, which is unavoidably
problematic. That is, use the data only where they exist. 
For this, one would have to Fourier transform $G_E(\mathbf q)$
into $\rho(r)$ and study its moments. This is probably the only
way of facing their unavoidable correlations.

\noindent (5) Show the correlated results as $(\langle r_p^2 \rangle,\langle r_p^3 \rangle_{(2)})$
contour plots for fixed acceptable values of $p$.

Such a procedure is very different from the usual. It may well
lead to significantly different conclusions. Doing this analysis --in contrast
to the simpler choice of verbally discussing it-- is well beyond the scope of this
paper.

\section{Discussion}

\begin{figure}[hbt!]
\hspace{-.4cm}
\includegraphics[width=0.47\textwidth]{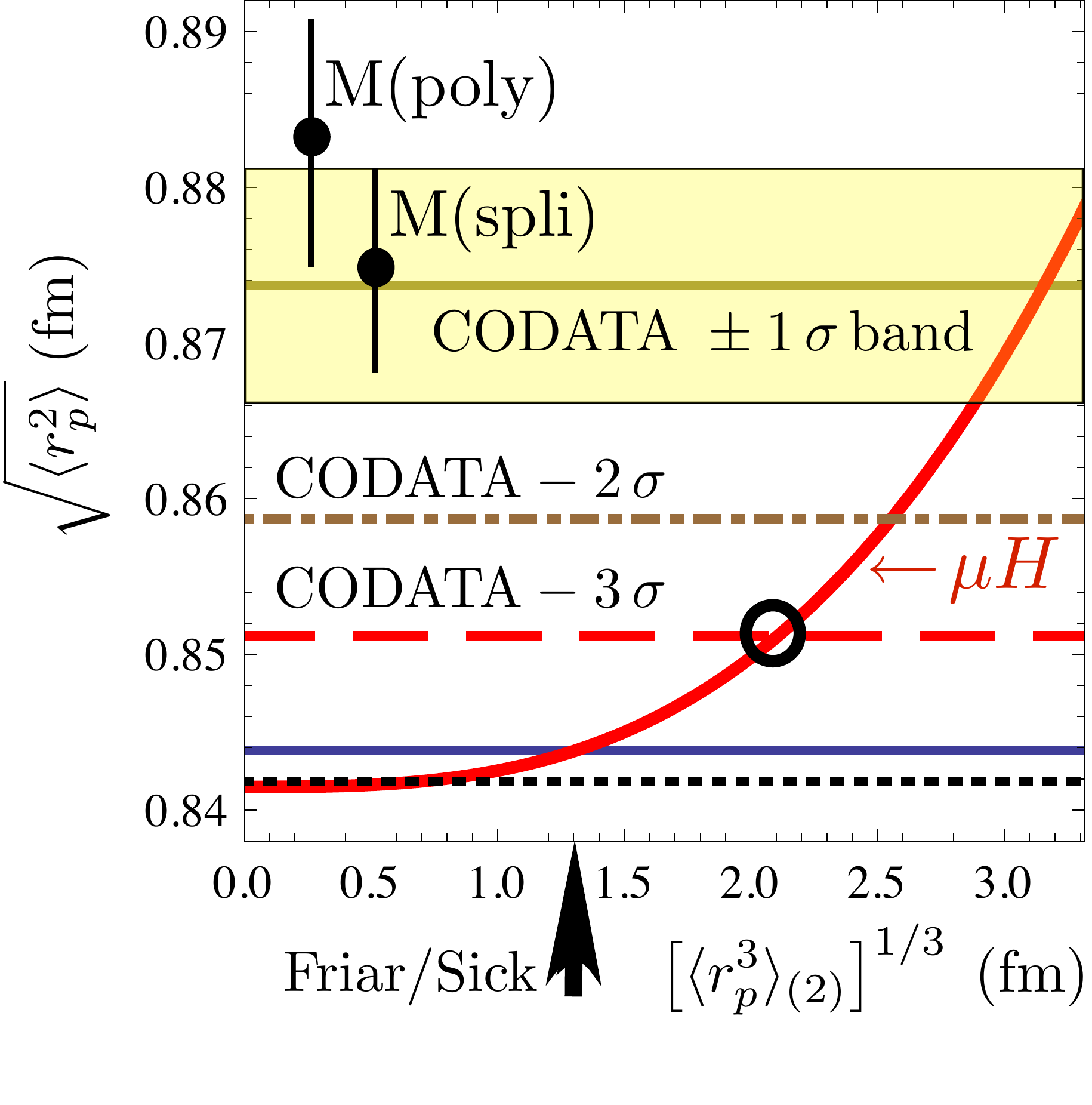}
\caption{Atomic data for the rms radius versus the third Zemach moment.
The ``$\mu H$" curve is the correlation between the rms radius and the third Zemach
moment implied by the observation of Eq.~(\ref{Lth}) and the
theoretical result of Eq.~(\ref{Lth2}).
Also shown are the polynomial and spline
results of  \cite{Mainz} (meant to be read as
horizontal bands) and the $\langle r_p^3 \rangle_{(2)}$ of \cite{FriarSick}.
  \label{fig:rPlot}}
\end{figure}

The result of Eq.~(\ref{rCODATA}) is not only based on measurements including
ordinary-hydrogen levels,
but also on the $ep$-scattering
result of Eq.~(\ref{rep}), discussed in Section V. The elimination of
this input from the CODATA fit to 78 (more or less) fundamental constants results in
\cite{CODATA}:
\begin{equation}
\sqrt{\langle r_p^2\rangle}(ep)=0.8737 \pm 0.0075\; \rm fm
\label{rephydrogen}
\end{equation}
This result is shown as the shaded band
in Fig.~\ref{fig:rPlot}, displaying the rms radius versus the cubic root of the
third Zemach moment. To facilitate the coming discussion, I have added
the $-2\sigma$ and $-3\sigma$ lines corresponding to Eq.~(\ref{rephydrogen}).
Also shown in the figure are the two results,
Eqs.~(\ref{rA1double}), of the Mainz  experiment \cite{Mainz}.

The lowest (dashed) line in Fig.~\ref{fig:rPlot} is Eq.~(\ref{rPohl}), from the $\mu H$
Lyman shift \cite{Pohl}. The continuous straight line
above the previous one takes into account the renormalization correction
of Eq.~(\ref{pole}). Make the same correction in Eq.~(\ref{Lth})
to obtain:
\begin{eqnarray}
\!\!\!\!\!\!L^{\rm th}\left[\langle r_p^2\rangle,\langle r_p^3 \rangle_{(2)}\right]&=&\nonumber\\
\!\!\!\!\!\!209.9779(49)&-&5.20123\, \langle r_p^2\rangle +0.00913 \,\langle r_p^3 \rangle_{(2)}
\label{Lth2}
\end{eqnarray}
and equate it to the observed value of Eq.~(\ref{Lexp}) to obtain the ``$\mu H$"
correlation, the continuous curve in Fig.~\ref{fig:rPlot}. 
This correlation --and not a figure for $r_p$-- is the outcome of the 
theoretical analysis of the measurement  \cite{Pohl}.

If the value of the
third Zemach moment was that of Eq.~(\ref{r3FS}),  indicated by an
arrow in Fig.~\ref{fig:rPlot}, the muonic- and ordinary-hydrogen results would
be more than $3\sigma$ away. If, instead,
$(\langle r_p^3 \rangle_{(2)})^{1/3}\sim 2.1$ fm the tension would diminish
to the $3\sigma$-level, marked by a circle in Fig.~\ref{fig:rPlot}
(the increase of the moment is more severe than it seems to be, since the
observable is not its cubic root).

The standard deviations of the previous paragraph are the ones pertinent
to a normal data distribution, for which $\pm 1$, $\pm 2$ and $\pm 3\sigma$
correspond to coverage probabilities of 68.27\%, 95.45\% and 99.73\%. But the data
of CODATA are not normally distributed, meaning that the bands of the same fixed
 probability are not the ones in Fig.~\ref{fig:rPlot}, and that the
conclusions of the previous paragraph should be correspondingly weakened.

More precisely, 9 out of 135 input data in \cite{CODATA}
 --related to the Watt balance, the lattice
spacing in various Silicon crystals, the molar volume of the same element and
the quotient of Plank's constant to the neutron mass)
have had their uncertainties increased by a multiplicative factor 1.5 \cite{CODATA}.
This choice helps in obtaining a fairly satisfactory overall $p$-value,  $p=0.221$, 
but it describes a hypothetical set of experiments, not the actual one.
Moreover, the question of the individual $p$-values of the experiments is not
reexamined in \cite{CODATA}. We have seen examples of how misleading
this omission may be.

\begin{figure}[hbt!]
\vspace{.5cm}
\hspace{-.4cm}
\includegraphics[width=0.45\textwidth]{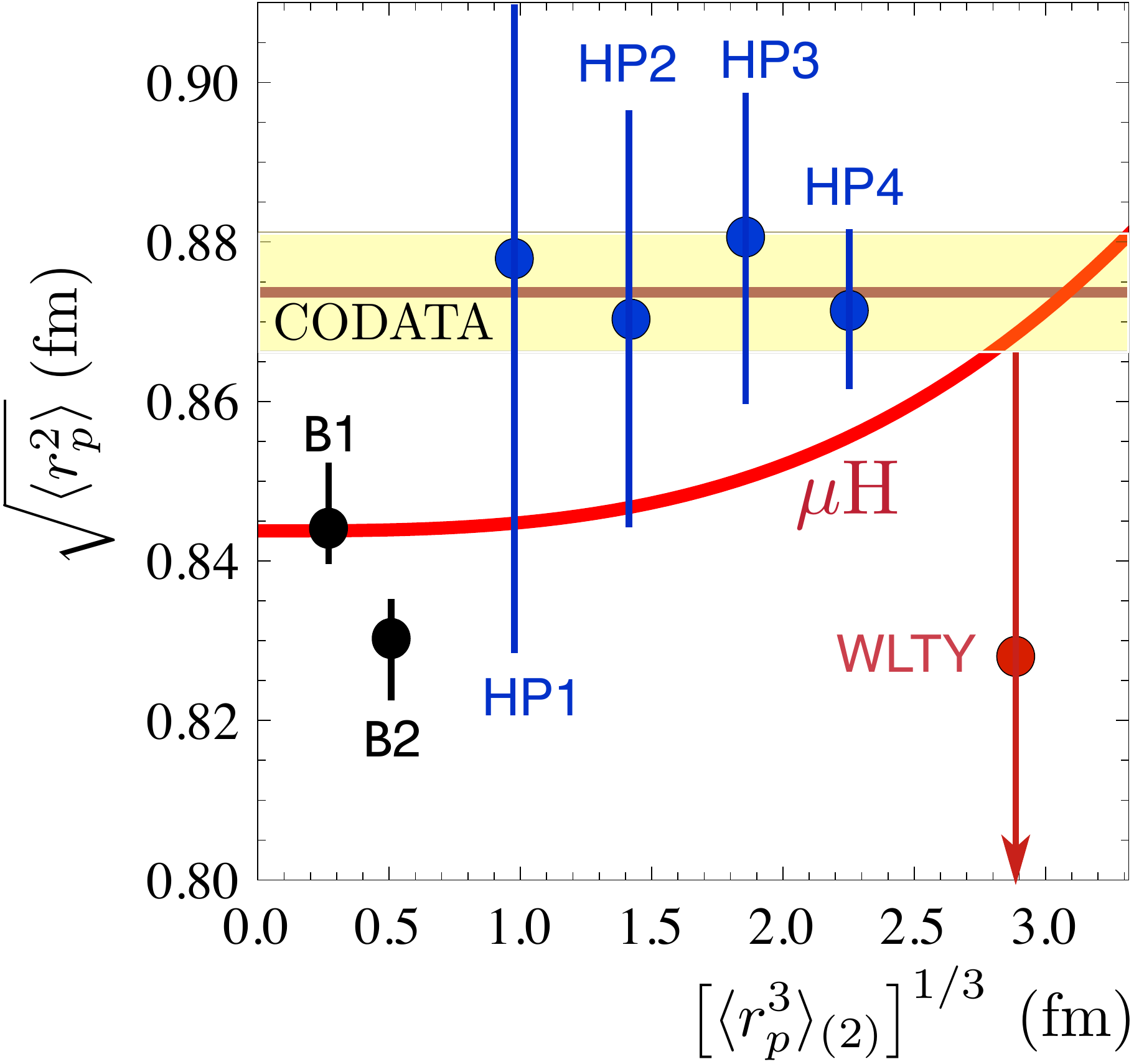}
\caption{The  CODATA result of 
Eq.~(\ref{rephydrogen}), the $\mu H$ correlation also shown in Fig.~\ref{fig:rPlot},
and recent ``theory-driven"
results for $\langle r_p^2\rangle$ (meant to be horizontal bands). 
B1, B2 cite \cite{DispersionRelations},
HP1 to HP4 cite \cite{HP} and WLTY quotes \cite{WLTY}.
  \label{fig:rPlotTheory}}
\end{figure}

A solidly-motivated approach to the extraction of $\langle r_p^2\rangle$
from the $ep$ scattering data is based on the use of the analytical properties of the
nucleon's form factors. Two recent examples are \cite{DispersionRelations, HP}
(it would be helpful to know the $p$-values of
these fits). 
Their $r_p$ outcomes are shown in Fig.~\ref{fig:rPlotTheory}.
The result of a first-principle lattice calculation \cite{WLTY}
is also shown. The spread of the theoretical results may be indicative of the
difficulty of reaching a consensus. Should the spread reflect a level of uncertainty,
there is no ``tension" between theory and observations.

\section{Conclusions}

We have seen that, to the precision required to analyse the recent 
muonic-hydrogen results, the vintage ``wave-function at the origin"
expression for the Lamb shift, Eq.~(\ref{oldr2shift}), is insufficiently
precise. In configuration space this is because the wave function
is not only probed at the origin, but up to distances at which details
of the proton charge distribution (beyond its mean square radius) 
become relevant. Thus the model-dependence of the
correction, see Eqs.~(\ref{pole},\ref{dipole}).

Translated into momentum space, the previous paragraph becomes
very familiar. A wave function at the ``origin", a $\delta(\vec 0)$,
corresponds to a uniform sampling of all momentum transfers, something
that an atom can hardly provide. The correction, a proper match of
renormalisation scales, can be phrased as a proton's running radius.
Not a surprising result: in a field theory all measured quantities are
scale-dependent.

The above correction to the analysis in \cite{Pohl}
reduces its disagreement with the $\langle r_p^2 \rangle$ CODATA
result from $\sim 5\,\sigma$ to at most $\sim 4\,\sigma$. The
``at most" is crucial, for the conclusion depends on the value adopted for the
third Zeemach moment. We have seen that its extraction 
from $ep$ scattering data is most questionable.
In Fig.~\ref{fig:rPlot} one can see that, even after the corrections I have
discussed, $(\langle r_p^3 \rangle_{(2)})^{1/3}\sim 3$ fm would be required
to have ordinary and muonic hydrogen precisely agree.
Even if a dipole form factor is only a very vague description of the data,
such a value feels unexpectedly large, part of the argument in
\cite{CM}.

The Lamb shift measurement provides a correlation between the two
relevant moments: the narrow curved domain labeled 
``$\mu H$" in Fig.~\ref{fig:rPlot}. 
It would be very helpful to extract the correlation dictated by $ep$
data, to be added as confidence-level contours to the figure, to decide
--with confidence-- what the empirical conclusion is. It may well be that
the $\mu H$ and $ep$ correlations have a sufficient overlap for the
question of data incompatibility to be moot. After all, also for $ep$
scattering, the two quantities plotted in  Fig.~\ref{fig:rPlot}
are obviously strongly correlated, a fact that has been totally ignored.

Similar inferences can be extracted from the comparison
of theory and data summarized in Fig.~\ref{fig:rPlotTheory}.
Currently none of these ``theory-driven"
 results are available in the form of two-dimensional 
$(\langle r_p^2 \rangle,\langle r_p^3 \rangle_{(2)})$
confidence-level plots. 
Even barring other putative limitations
of current theory or experimental analyses, the most extreme views
on the subject at hand \cite{rpADR, Flowers} seem to have been largely
exaggerated.

\subsection*{Acknowledgments}
I am grateful to Shelly Glashow, Maurizio Pierini, Chris Rogan
and Maria Spiropulu for their comments and to Pietro Slavich
for finding an error in \cite{Yo1}.



\begin{thebibliography}{999}


\bibitem{Pohl}
R.~Pohl,  {\it et al.} Nature {466}  (2010) 213.
R. Pohl, et al., Supplementary Information, doi: 10.1038/nature09250.
\bibitem{CODATA}
P.~J.~Mohr, B.~N.~Taylor  \& D.~B.~Newell,  Rev. Mod. Phys. {80} (2008) 633.
\bibitem{rpADR}
A. De R\'ujula, Phys. Letts. B 693 (2010) 555.
\bibitem{Flowers}
J. Flowers, Nature 466 (2010) 195.
\bibitem{Mainz}
Bernauer et al. arXiv:1007.5076v2 (2010).
\bibitem{HTheory}
M.~I.~Eides, H.~Grotch \& V.~A.~Shelyuto,  Springer
Tracts in Mod.~Phys.  {222}  (Springer, Berlin Heidelberg,
2007).
S.~G.~Karshenboim, Phys. Rep. {422} (2005) 1.
\bibitem{Hexps}
M.~Niering,   Phys. Rev. Lett. {84} (2000) 5496. 
B.~de Beauvoir,   Eur. Phys. J. {D 12} (2000) 61.
C.~Schwob Phys. Rev. Lett. {82} (1999) 4960.
\bibitem{Sick}
I.~Sick,  Phys. Lett. {B 576} (2003) 62.
\bibitem{Sick2}
P.~G.~Blunden \& I.~Sick,  Phys. Rev. {C 72}  (2005) 057601.
\bibitem{LymanTH}
S.~G.~Karshenboim,   Phys. Rep. {422} (2005) 1.
K. Pachucki,
Phys. Rev. {  A60} (1999) 3593.
 E.~Borie,  Phys. Rev. {A 71}  (2005) 032508.
A.~P.~Martynenko,  Phys. Rev. {A 71} (2005) 022506. 
A.~P.~Martynenko, 
 Phys. At. Nucl. {71} (2008) 125.\
 K.~Pachucki \& U.~D.~Jentschura,   Phys. Rev. Lett. {91}  (2003) 113005.
 \bibitem{FriarSick}
J.~L.~Friar \& I.~Sick  Phys. Rev. {  A 72}  (2005) 040502(R).
\bibitem{A}
A. De R\'ujula, Phys. Rev. Lett. 32 (1974) 1143.
The argumentation in this paper was strengthened in
A. De R\'ujula, H. Georgi \& H.D. Politzer, Ann. Phys. (N.Y.) 103 (1977) 315;
Phys. Lett. 64B (1976) 428. For a lighter discussion, see
A. De R\'ujula in {\it 50 Years of Yang-Mills Theory},
edited by G. 't Hooft. World Scientific Publishing Co. Singapore, 2005,
page 401.
\bibitem{GT}
D.~J. Gross \& S.~B. Treiman, Phys. Rev. Lett. 32 (1974) 1145.
\bibitem{PGW}
H.~D. Politzer, Phys. Rev. Lett. 26 (1973) 1346.
D. Gross \& F. Wilczek, Phys. Rev. Lett. 26 (1973) 1343.
G. 't~Hooft, unpublished.

\bibitem{Simon}
G.G. Simon, et al., Nucl. Phys. A 364 (1981) 285.
\bibitem{CM}
I.~C.~Cl$\ddot{\rm o}$et \& G.~A.~Miller, arXiv:1008.4345v3.
\bibitem{DispersionRelations}
M.~A.~Belushkin, H.~W.~Hammer \& U.~G.~Meissner, Phys. Rev. {  C 75} (2007)  035202.
\bibitem{HP}
R.~G.~Hill \& G.~Paz, arXiv:1008.4619v1.
\bibitem{WLTY}
P.~Wang et al., Phys. Rev. D 79 094001 (2009).
\bibitem{Yo1}
A.~De R\'ujula, arXiv:1010.3421.









\end{thebibliography}
\end{document}